\documentclass[prd,aps,floats,epsfig,twocolumn]{revtex4}
\usepackage{graphicx}
\def\beq{\begin{equation}}
\def\eeq{\end{equation}}
\def\bea{\begin{eqnarray}}
\def\eea{\end{eqnarray}}
\def\p0{\phi_{0}}

\def\eg{{\it e.g. }}
\def\ie{{\it i.e. }}

\newcommand{\Nature}{{\it Nature\,}}

\newcommand{\ApJ}{{\it Astrophys. J.\,}}
\newcommand{\ApJL}{{\it Astrophys. J. Lett.\, }}

\newcommand{\PR}{{\it Phys. Rev.\,}}

\newcommand{\PL}{{\it Phys. Lett.\,}}

\newcommand{\etal}{{\it et al.}}

\begin{document}

%\twocolumn[\hsize\textwidth\columnwidth\hsize\csname@twocolumnfalse\endcsname
%\date{\today}
\title{Are Chaplygin gases serious contenders to the dark energy throne? }
\author{Rachel Bean and Olivier Dor$\acute{e}$ }
\address{Department of Astrophysical Sciences, Princeton University,
  Peyton Hall - Ivy Lane, Princeton, NJ08544-1001, USA\\
  rbean@astro.princeton.edu, olivier@astro.princeton.edu } 

\begin{abstract}
We study the implications on both background and perturbation
evolution of introducing a Chaplygin gas  component in the universe's
ingredients. We perform likelihood  analyses using wide-ranging, SN1a,
CMB and large scale structure  observations to assess whether such a
component could be a genuine alternative to a cosmological constant,
$\Lambda$. We find that the current data favors behavior in an adiabatic Chaplygin Gas that is akin to a cosmological constant.

\end{abstract}
\maketitle
%\pacs{PACS Numbers: }
%]

\renewcommand{\thefootnote}{\arabic{footnote}}
\setcounter{footnote}{0}

\section{Introduction}
%============================================================

Supernovae observations \cite{super1,superP} first indicated that the universe's
expansion has started to accelerate during recent cosmological
times. This, and further observations, \eg of the Cosmological Microwave
Background (CMB) or Large Scale Structures (LSS) , suggest that the energy
density of the universe is dominated by a dark energy component, with a negative pressure,
driving the acceleration. One of the substantive goals for
cosmology, and for fundamental physics, is ascertaining the nature
of this dark energy. Maybe the most attractive option would be a
cosmological constant, $\Lambda$, however there are infamous
fine-tuning and coincidence problems associated with explaining
why $\Lambda$ should have today's energy scale. These problems
have lead to a wealth of dynamical, scalar, dark energy
(``quintessence") models being proposed as alternatives to
$\Lambda$ (see \cite{Peebles02} for a good review).  Even in these
cases, however, explaining why our epoch should be so crucial in
triggering the acceleration still requires fine-tuning.

A concurrent problem is the nature of the non-baryonic, clumping
dark matter component required in the standard model to give Large
Scale Structure predictions consistent with observations.

Recently an alternative matter candidate, a Generalized Chaplygin
Gas (GCG), has been proposed as a potential `hybrid' solution to both
the dark energy and dark matter problems. The GCG can be seen to
evolve in a wide range of contexts, for example from supersymmetry, tachyon cosmologies \cite{Gibbons03} and brane cosmologies \cite{Bilic01}.  A recent letter \cite{Teg02} dealt with the implications for the matter power spectrum in the absence of CDM and effectively ruled out the GCG as a CDM substitute. 

In this paper we investigate the strength of the GCG as a dark energy candidate.
Although there have been a number of papers discussing various
aspects of GCG behavior (\cite{Kamen01}-\cite{Gonz02}) there has
not been, as yet, a full analysis of the constraints that can be
placed on such models from the wide range of complementary data
sets currently available. This is necessary if such exotic matter
types are to be considered as serious alternatives to the
$\Lambda$-CDM scenario.

In section \ref{sec2} we review the background evolution of the
GCG  and discuss the implications for supernovae (SN1a) observations.
Although such constraints are important, a wide range of proposed
theories can generate the required expansion profile (see
\cite{Peebles02} for dark energy theories and, for example,
\cite{Deffayet02} for an alternative to dark energy). In order to
better discriminate between theories, perturbation dependent
observables must be taken into consideration. In section
\ref{sec3} we extend our discussion to perturbations in the
Chaplygin gas and discuss the implications for
structure formation in the presence of 
an adiabatic Chaplygin fluid. In section \ref{sec4} we
consider the effects on radiation perturbations and the CMB spectrum. In section
\ref{sec5} we present the main results of the paper, likelihood
analyses for a CDM+GCG+baryon
universe. We include the option of a pure GCG + baryon scenario ($\Omega_c=0$) for completeness. We obtain a clear indication of the strength of the GCG
model when compared to CDM and $\Lambda$. In section \ref{sec6}
we summarize our findings and assess the true potential of
Chaplygin gases as a dark energy contender.

\section{Background evolution}
%============================================================
\label{sec2}

The Generalized Chaplygin models can be characterized by three
parameters: $w_0$, $\alpha$ and $\Omega_{ch}^{0}$. The equation of state nowadays $w(a=1)=-|w_{0}|$
 and the index $\alpha$ specify the equation of state
evolution,
\beq 
p=-{|w_{0}|\Omega_{ch}^{0}\rho_{0}\over \rho^{\alpha}}.
\label{peq} 
\eeq
where $\rho_{0}=3H_{0}^{2}$  $(8\pi G=1)$ is the
total energy density today.
The energy conservation equation, $\dot{\rho}+3{\cal H}(1+w)\rho=0$,
admits a solution for $\rho(a)$ specified by $|w_{0}|$, $\alpha$ and
the fractional energy density today, $\Omega_{ch}^{0}$,
\beq
\rho(a)=\Omega_{ch}^{0}\rho_{0}\left[|w_{0}|+{(1-|w_{0}|)\over
 a^{3(1+\alpha)}}\right]^{{1\over1+\alpha}}
\label{rhoeq} \eeq . The equation of state then evolves as,
\beq w(a)=-{|w_{0}|\over \left[|w_{0}|+{(1-|w_{0}|)\over
a^{3(1+\alpha)}}\right]}. \label{weq}\eeq 
 At early times the GCG's
equation of state tends to zero, mimicing CDM . The
value of $\alpha$ determines the redshift of transition between
the two asymptotic behaviors; the greater the value of
$\alpha$ the lower the transition redshift. At early times, the total
amount of matter with $w\sim 0$ reaches an asymptotic value

\beq \Omega_{m,eff}^{0}=\Omega_{m}^{0}+\Omega_{ch}^{0}\left(
1-|w_{0}|\right)^{{1\over 1+\alpha}} \label{rhoearly}\eeq where
$\Omega_{m}$ is the baryonic + CDM density fraction. Note that
  the unique ability of the GCG to account for both the dark energy like behavior
at late times and for ordinary dark matter at early times motivated
the original studies of this particular equation of state.

In previous discussions $\alpha$ has often been assigned a
positive value in the range $0< \alpha <1$, in order to be
consistent with various higher dimensional theories that can
produce a perfect fluid stress-energy tensor satisfying the
criterion in (\ref{peq}) (see for example \cite{Podolsky02}). In
our study, we extend the range of values of $\alpha$ considered to
$-1<\alpha<\infty$ in order to obtain a broader assessment of
whether Chaplygin gases could be a viable alternative to the
standard model. 

For $\alpha$=0 the background evolution of the Chaplygin
gas is identical to a $\Lambda$-CDM model with
$\Omega_{\Lambda,eff}=\Omega_{ch}|w_{0}|$, and
$\Omega_{m,eff}=\Omega_{m}+\Omega_{ch}(1-|w_{0}|)$. Furthermore, as is
visible in equations (\ref{rhoeq}) and (\ref{weq}), when $|w_0|$ tends to
1, the GCG component tends to evolve as a cosmological constant,
irrespective of the value of $\alpha$. Note that there is no
analogous quintessence like behavior (with $w_0 \ne -1$), thus we are
only comparing GCG to theories including $\Lambda$.

%----------------------------------------------------------
\begin{figure}
\centerline{\includegraphics[width=8cm]{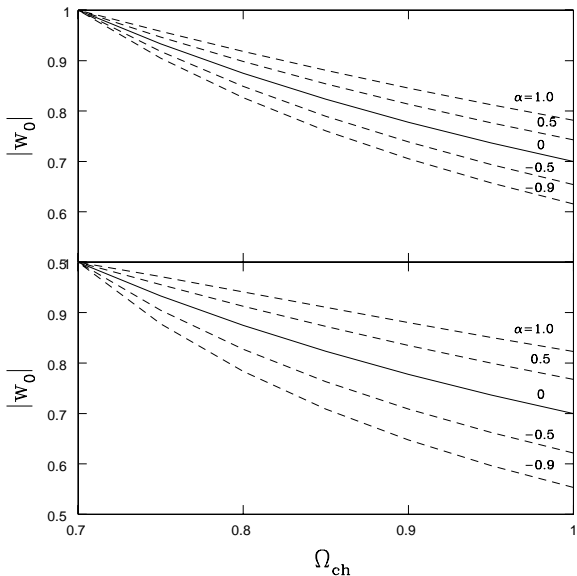}}
\caption{Contours in $\Omega_{ch} - |w_{0}|$ space with the same
  luminosity distance as a fiducial model $\Lambda$-CDM model with
  $\Omega_{\Lambda}=0.7$ at $z=0.5$ (top) and $z=1$ (bottom). For
  $\alpha=0$ (full line) the luminosity distance curve is identical to
  that of the fiducial model at all redshifts.} \label{fig1} 
\end{figure}
%----------------------------------------------------------

The SN1a observations measure the apparent magnitude, $m(z)$,
related to the luminosity distance, $d_{L}(z)$ via, \bea
m(z)&=&{\cal M}+5\log{d_{L}(z)}+25, \\  d_{L}&=&(1+z)\int_{0}^{z}
{dz'\over H(z')} \eea
where $\cal{M}$ is the absolute bolumetric magnitude and $d_{L}$
is measured in Mpc. It is easy to see that, because the background
evolution (through $H$) wholly determines 
luminosity distance predictions, the degeneracy between a GCG with $\alpha=0$ and $\Lambda$-CDM
will allow the Chaplygin gas to fit the SN1a data well.
Indeed the degeneracy also stretches to $\alpha\ne 0$ when one considers luminosity distance at a specific redshift. In figure \ref{fig1} we show Chaplygin models with degenerate
luminosity distances with a fiducial $\Lambda$-CDM model with
$\Omega_{\Lambda}$=0.7, at $z=0.5$ and 1.0. 
This degeneracy, however, implies that the SN1a observations cannot be a strong
discriminant between the GCG and $\Lambda$; we must look to
alternative, perturbation-dependent observations to test the validity
of the GCG models.

\section{Chaplygin gas perturbations}
%============================================================
\label{sec3}

We treat the Chaplygin gas as a perfect fluid made up of
effectively massless particles interacting with the rest of matter
purely through gravity. We assume purely adiabatic contributions
to the perturbations so that the speed of sound for the fluid is
\beq c_{s}^{2}={\delta
p \over \delta \rho}={\dot{p} \over
\dot{\rho}}=-w\alpha \label{cs2}\eeq
and the time variation of $w$ is
\beq
\dot{w}=-3{\cal H}\left(1+w\right)\left(c_{s}^{2}-w\right)=3{\cal H}w(1+w)(\alpha+1)
\eeq
where derivatives are with respect to conformal time (d/d$\tau$), and $a{\cal H}=da/d\tau$.
%
%----------------------------------------------------------
\begin{figure}
\centerline{\includegraphics[width=8cm]{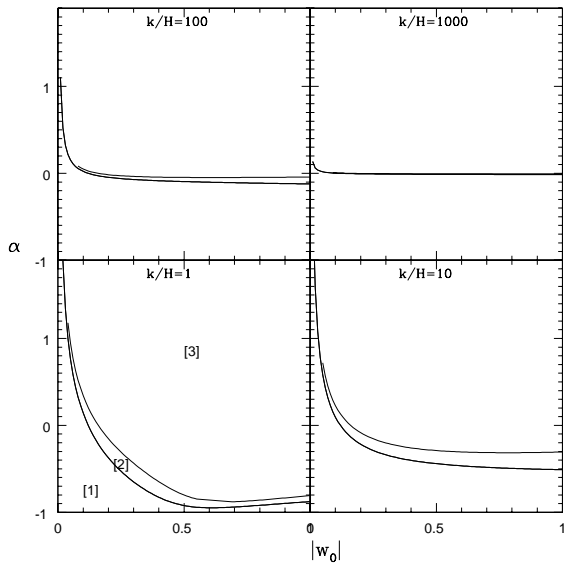}}
\caption{Late-time $\delta$ evolution as a function of $\alpha$
and $|w_{0}|$  for 4 scales $k/{\cal
H}$ = 1,10,100,1000 (${\cal H}\approx 3000h
Mpc^{-1}, h=0.65$). Regions [1] and [2] undergo power law growth and
decay respectively and [3] undergoes oscillatory decay. } \label{fig2} 
\end{figure}
%----------------------------------------------------------

In the synchronous gauge and following the approach and notations
of Ma and Bertschinger \cite{Ma95}, we can write down the
evolution equations for the density and velocity divergence
perturbations, $\delta$ and $\theta$, using the conservation of
energy momentum tensor $T^{\mu}_{\nu; \mu}=0$,
\bea
\label{deltaeq}\dot{\delta}&=&-\left(1+w\right)\left(\theta+{\dot{h}\over
  2}\right)-3{\cal H}\left(c_{s}^{2}-w\right)\delta, \\ 
\label{thetaeq}\dot{\theta}&=&-{\cal H}\left(1-3c_{s}^{2}\right)\theta+{c_{s}^{2}\over
(1+w)}k^{2}\delta-k^{2}\sigma. \eea
The fluid is highly non-relativistic and therefore we assume the
shear perturbation $\sigma=0$.

At early times, when the Chaplygin gas has $w\approx 0$, the GCG
perturbations evolve like those of ordinary dust with
$\dot{\theta}=\theta=0$, and $\dot{\delta}=-\dot{h}/ 2$. In the
radiation era $\delta(a)\propto a^{2}$ , while $\delta \propto a$
in the early GCG dominated era. At later times, when the GCG's equation
of state starts to decrease, the perturbations stray drastically
from this dust-like evolution.

We can understand the late-time behavior more clearly if we
evaluate the second order differential equation for $\delta$. By
differentiating equation (\ref{deltaeq}) with respect to time we
find, as outlined in an appendix (section \ref{app}), for a general,
shearless, fluid,
\bea 
\label{deltaeq2}
\ddot{\delta}&+&\left[1+6(c_{s}^{2}-w)\right]{\cal H}\dot{\delta}
\nonumber \\ &+&\left[9{\cal H}^{2}(c_{s}^{2}-w)^{2} +3{\cal
H}(\dot{c_{s}^{2}}-\dot{w})+3{\ddot{a}\over
a}(c_{s}^{2}-w)+c_{s}^{2}k^{2}\right]\delta \nonumber \\ &&
=-3c_{s}^{2}(1+w){\cal H}\theta+{a^{2}\over 2}(1+w)(3\delta P
+\delta \rho). 
\eea
%----------------------------------------------------------
\begin{figure}
\centerline{\includegraphics[width=8cm]{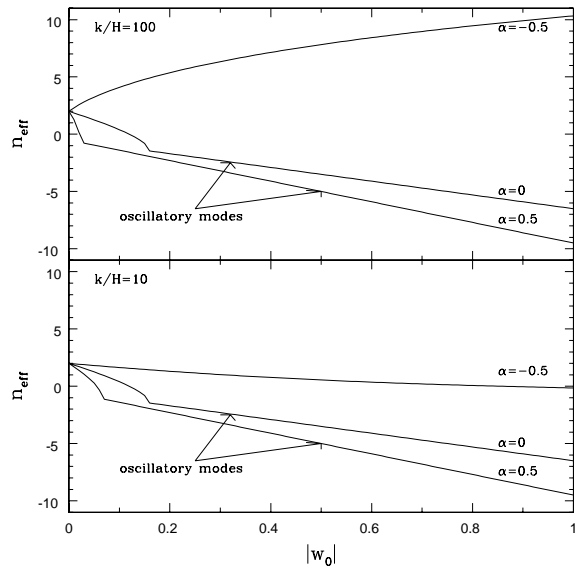}} \caption{Late
time evolution envelope for $\delta$. The power,
$n_{eff}=\tau\dot{\delta}/\delta$, is plotted for two length
scales $k/{\cal H}$ = 10  and 100. In the
oscillatory regime $n_{eff}$ for the bounding envelope is plotted.
 } \label{fig3}
\end{figure}
%----------------------------------------------------------

Numerical integration shows that the coupling to $\theta$ in
equation (\ref{deltaeq2}) is subdominant for all scales that we
are interested. For the Chaplygin gas,
\bea\label{deltaeq3}
\ddot{\delta}&+&A{\cal H}\dot{\delta}+B{\cal H}^{2}\delta-{3{\cal
    H}^{2}\over 2}(1+w)\Omega_{c}\delta_{c}=0, \\  
A&=&1-6w(\alpha+1), \\ 
B&=&-{3{\cal H}^{2}\over
2}\left\{\Omega_{ch}+\left[7+\Omega_{ch}+(13-3\Omega_{ch})\alpha+6\alpha^{2}\right]w
\right.\nonumber \\ &&\left.-3\Omega_{ch}(1+2\alpha)w^{2}
+{2\alpha w\over 3}\left({k\over{\cal H}}\right)^{2}\right\},
\eea
where the subscript $c$ refers to cold dark matter.

For $w=0$, (\ref{deltaeq3}) reduces to the expected, scale
independent, CDM evolution with $\delta\propto \tau^{2}$. For
$w\neq 0$ we retain scale independence if $\alpha=0$ and just get
suppression of density perturbations. However for $\alpha\neq 0$ the perturbation
evolution becomes scale dependent with the $(k/{\cal H})^{2}$ term
dominating the others for scales greater than a characteristic
scale \beq k_{*}^{2}={{\cal H}^{2}\over |\alpha w|}.\eeq There are
3 possible solution types, for  $k\gg k_{*}$,
\beq\begin{array}{cll}
\textrm{[1]}: &\textrm{growing\ mode},& \alpha<0 ,\\
\textrm{[2]}: &\textrm{decaying\ mode},&\alpha\sim 0 ,\\
\textrm{[3]}: &\textrm{oscillatory decay},&\alpha>0 .\label{odetab}
\end{array}
\eeq
In figure \ref{fig2} we show the asymptotic behavior at late times
(taking $w\approx -|w_{0}|$) as one increases the scale $k/{\cal
H}$.  In figure \ref{fig3} we show the
associated scaling of $\delta$, plotting $n_{eff}=\tau\dot{\delta}/\delta$. For $|w_{0}|\neq 0$, 
and  $\alpha>0 (<0) $ GCG perturbations are suppressed (promoted) in comparison to those for a
$\Lambda$-CDM model.

\begin{figure}
\centerline{\includegraphics[width=8cm]{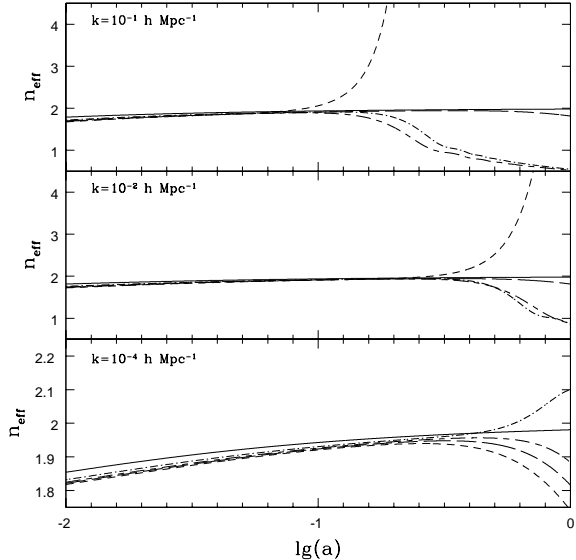}} \caption{
Evolution of  $n_{eff,c}=\tau\dot{\delta_{c}}/\delta_{c}$ for
$|w_{0}|=0.5$ and $\alpha$=-0.1 (short
dash), 0 (long dash), 0.1 (long-short dash), 0.5 (dot dash) in
comparison to pure CDM model (full line).  Three
length scales $k = 10^{-4},10^{-2},10^{-1} h Mpc^{-1}$ are considered. } \label{fig4}
\end{figure}

The GCG also has an effect on the the CDM perturbations through the relation:
\bea
\label{cdmeq} 
\ddot{\delta}_{c}&+&{\cal H}\dot{\delta}_{c}-{3{\cal H}^{2}\over 2}\left[\Omega_{c}\delta_{c}+(1-3\alpha
w)\Omega_{ch}\delta\right]=0 .\eea 
If $|w_{0}|\ne 0$ and $\alpha>0 (<0)$ the GCG drives
suppression (growth) in $\delta_{c}$. In
figure \ref{fig4} we show the power law evolution of
$\delta_{c}$, plotting $n_{eff,c}=\tau\dot{\delta}_{c}/\delta_{c}$ for 3 scales as one
varies $\alpha$. The strong growth in $\delta_{c}$ effectively rules
out a GCG with $\alpha<0$ as a dark energy candidate.

Note that the $\alpha$=0 degeneracy present in the background
evolution is not found in the perturbations. The matter power
spectrum and CMB spectrum will therefore be better discriminators
between $\Lambda$ and the GCG than SN1a. 
%----------------------------------------------------------
%----------------------------------------------------------

\section{Implications for temperature anisotropies}
%============================================================
\label{sec4} 

A Chaplygin gas matter component would change the
temperature anisotropy specturm in a number of ways; altering the
late-time ISW effect,  the peak positions and relative heights. 

The GCG's late-time evolution will alter the evolution of the
gravitational potential the  CMB photons pass through to reach us, inducing an ISW effect. 
Following \cite{Seljak96} and again using the terminology of
\cite{Ma95}, the ISW temperature anisotropy is given by a source,
\bea 
S_{ISW}&\propto& -\dot{\Psi}+\dot{\Phi} \\ 
&\propto& {d \over d\tau}\left[-{3\over
2}(\rho+P)a^{2}\sigma-a^{2}\delta\rho-3{\cal H}a^{2}(\rho+P){\theta\over
k^{2}}\right]\nonumber
\eea 
where $\Phi$ and $\Psi$ are the Bardeen variables \cite{Bardeen}.

%----------------------------------------------------------
\begin{figure}
\centerline{\includegraphics[width=8cm]{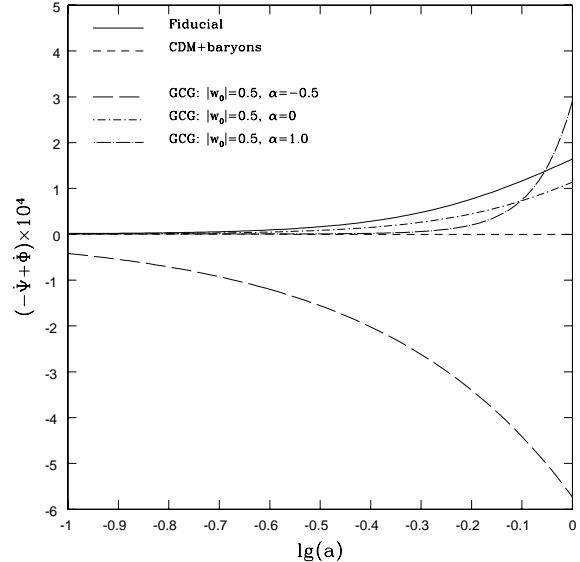}} \caption{Large
scale ISW effect for GCG with $\alpha$=-0.5 (long dash), 0 (short
dash-dot) and 1 (long dash -dot) compared against a CDM+baryon (short
dash) and `fiducial' $\Omega_{\Lambda}=0.7,\Omega_{m}=0.3$ (full
line) models.} 
\label{fig5}
\end{figure}

%----------------------------------------------------------
At late times the shear, $\sigma$, is negligible and it is the
density perturbation that drives the ISW effect. \bea {d\over
d\tau}\left[-{a^{2}\delta\rho}\right]
&\sim& a^{2}\left[(1+3w)-{n_{eff}\over p}\right]{{\cal
H}\rho\delta}\label{isw} \eea where $n_{eff}$ is the power law
index and $p= {\cal H}\tau$ as described in section \ref{sec3}.
Equation (\ref{isw}) shows why in a standard CDM scenario, with $w=0$
and $n_{eff}$ and  $p$ both $\approx 2$, there is no appreciable ISW effect.
In section \ref{sec3} however we saw for $\alpha<0$ that
$n_{eff}>2$ giving a negative ISW effect, while for $\alpha>0$,
$n_{eff}<2$ producing an increase in the ISW temperature
anisotropy. These effects are shown in figure \ref{fig5}.

The position of the first peak will be altered through adjustments to
the sound horizon, $r_{shor}$, and angular diameter distance 
at the last scattering surface, $d_{A}$. The position of the first peak in multipole space is given by \beq
l_{A}={\pi d_{A}(z_{rec})\over r_{shor}(z_{rec})}\eeq 
with
\bea d_{A}&=&\tau_{0}-\tau_{rec} \\&\approx&{1\over
  H_{0}}\int^{1}_{a_{rec}}{da\over
  \left[\Omega_{m}^{0}a+\Omega_{ch}^{0}a^{4}(|w_{0}|-{1-|w_{0}|\over
      a^{3(1+\alpha)}})^{1\over 1+\alpha}\right]^{1\over 2}}
\nonumber\\ 
r_{shor}&=&\int_{0}^{\tau_{rec}}c_{s}^{\gamma b}d\tau \\ 
&\approx&{1\over H_{0}
\sqrt{\Omega_{m,eff}^{0}}}\int_{0}^{a_{rec}}{c_{s}^{\gamma b}da\over a^{{1\over 2}}}
\nonumber\eea where $\Omega_{m,eff}^{0}$ is defined in equation
(\ref{rhoearly}) and $c_{s}^{\gamma b}$ is the speed of sound for the radiation-baryon system, not to
be confused with $c_{s}$ for the Chaplygin gas (at this time
the Chaplygin gas is behaving like dust and has $c_{s}^{2}=0$).
For fixed $\omega_{b}=\Omega_{b}h^{2}$ and $\omega_{c}=\Omega_{c}h^{2}
(h=H_{0}/100)$, $l_{A}$ increases as one increases $\alpha$ or $|w_{0}|$. The position of the first peak would be the same
for scenarios with the same value of
$\sqrt{\Omega_{m,eff}^{0}}d_{A}$. 

%----------------------------------------------------------
\begin{figure}
\centerline{\includegraphics[width=8cm]{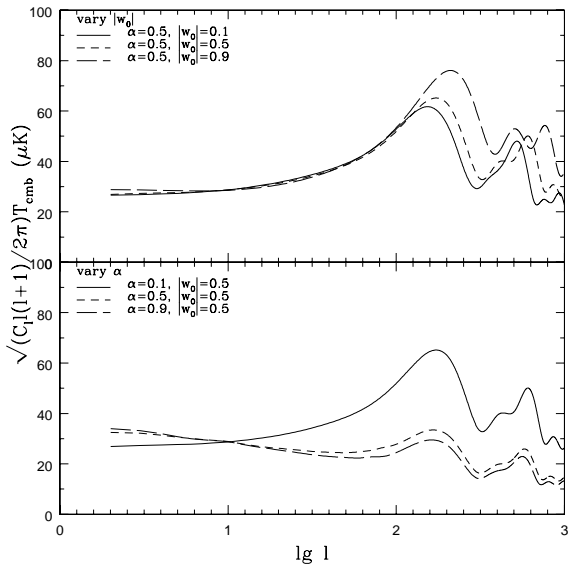}}
\caption{Comparison of CMB power spectra, normalized to COBE at
  $l=10$, with varying $\alpha$ and $|w_{0}|$, keeping all other
  relevant quantities fixed. The behavior of peak heights and
  positions is discussed in the text.}  
\label{fig6}
\end{figure}
%----------------------------------------------------------

Because the Chaplygin gas mimics matter at early times there is no
simple degeneracy, governed by the peak positions, as there is for
quintessence models  (see for example \cite{Bean01}). The peak
heights, when compared to the low $l$ `plateau', depend upon $\alpha$
and $|w_{0}|$ through their influence on the ISW effect, the horizon
scale at matter-radiation equality, $l_{eq}$, (through
$\Omega_{m,eff}^{0}h^{2}$) and the depth of the potential well at last
scattering (also through $\Omega_{m,eff}^{0}h^{2}$). 

We follow the phenomenological discussion in \cite{Hu01}, to predict
how the peak heights will alter for fixed $\omega_{b}$ and
$\omega_{c}$. Increasing $\alpha$, increases $\Omega_{m,eff}^0$, so that
matter-radiation equality happens earlier, increasing $l_{eq}$ and
curtailing the driving effect that the decay of the gravitational
potential has on $\delta_{\gamma}$ oscillations during the
radiation era. This lowers the height of the first peak, a decrease
which is compounded by the raising of the plateau from the ISW
effect. An earlier matter-radiation equality also decreases the depth
of the potential well at last scattering, which combined with the
reduction in radiation driving, increases the height of the third peak
in comparison to the first and second ones. 

As one increases $|w_{0}|$ one decreases $\Omega^{0}_{m,eff}$, lowering $l_{eq}$, and increasing
the height of the first peak. This is tempered, however, by the
increase in plateau height from the ISW effect. Reducing $l_{eq}$ acts
to decrease the height of the third peak in comparison to the second
and first ones. These behaviors are confirmed by the full analysis, as
is shown in figure \ref{fig6}. 

The multitudinous effects that the GCG has on the CMB spectrum make
comparison with CMB observations a strong test for the GCG models as
will be seen below.   

%----------------------------------------------------------
\begin{figure*}
\centering{
\hbox{\includegraphics[width=8cm]{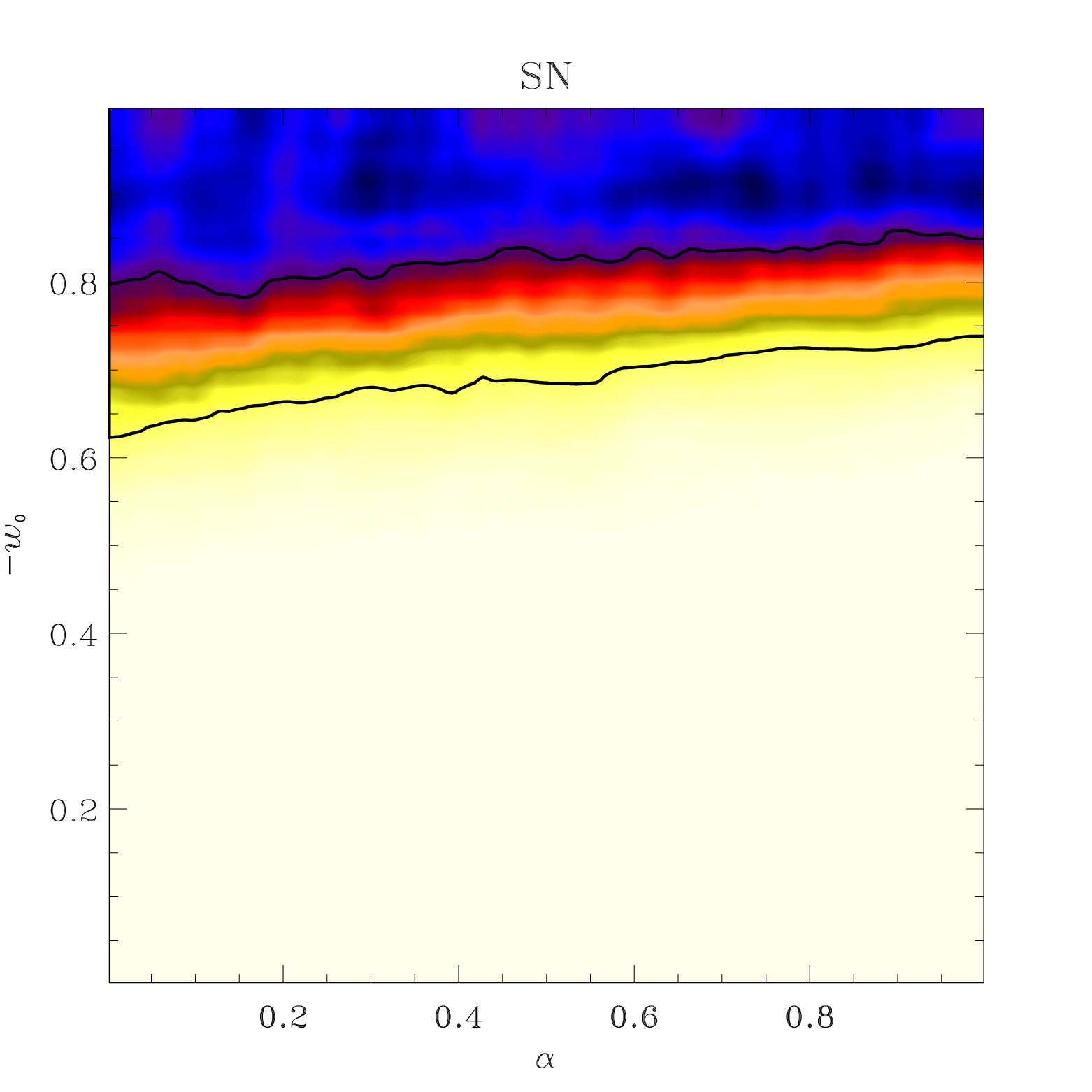}\includegraphics[width=8cm]{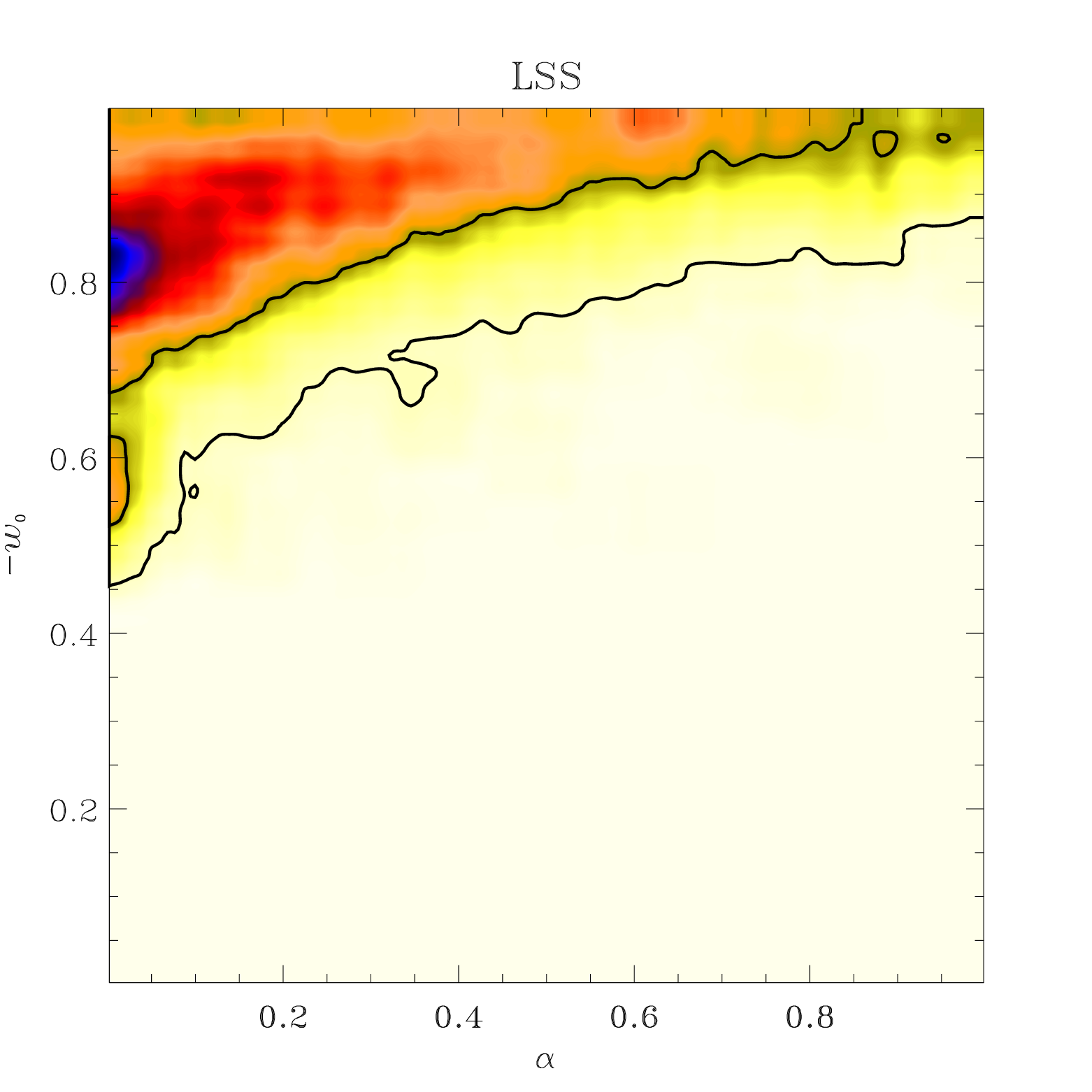}}
\hbox{\includegraphics[width=8cm]{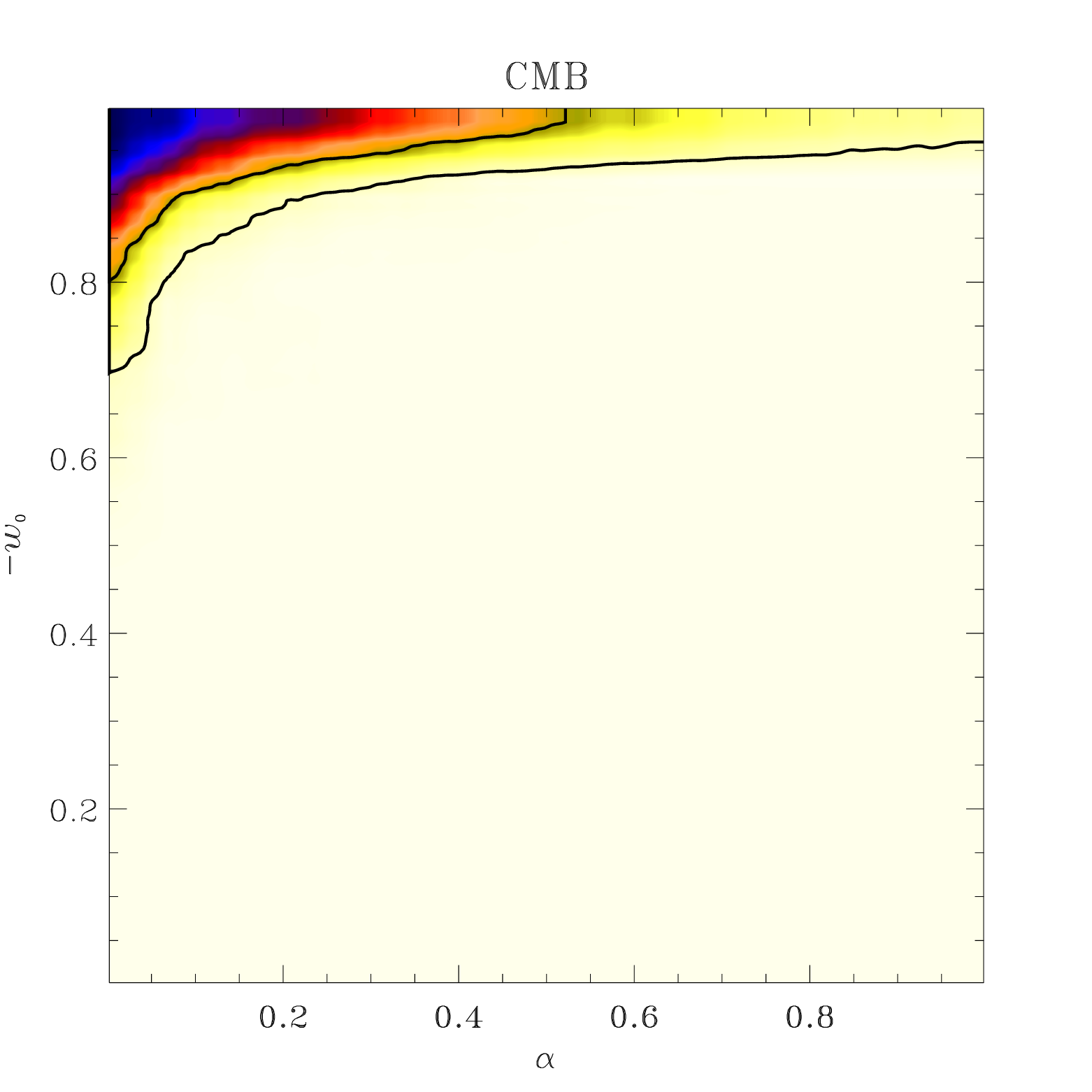}\includegraphics[width=8cm]{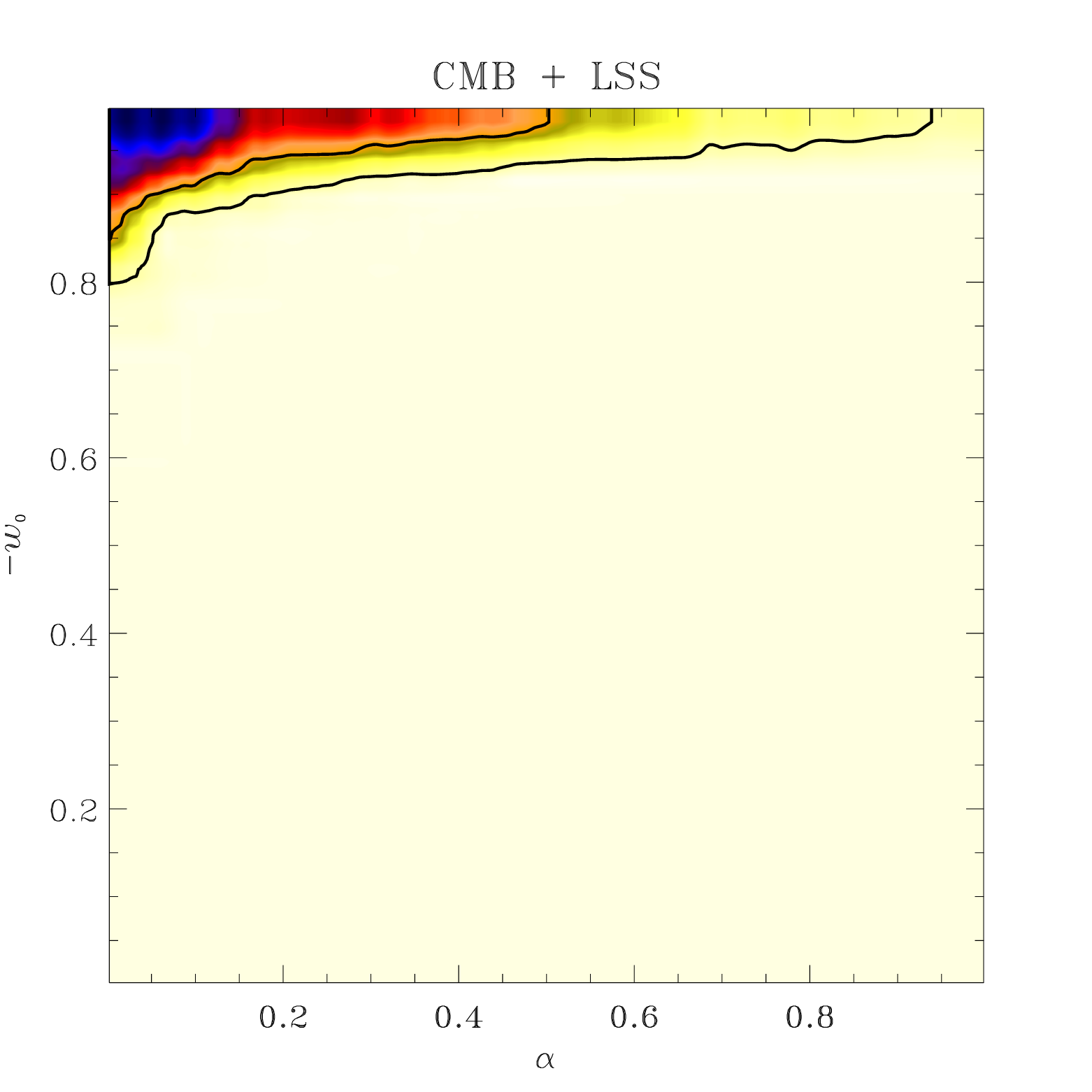}}
}
\caption{Joint posterior of the $\alpha$ and $-w_0$ parameters
  considering only SN1a data (top-left panel), LSS data (top-right
  panel), CMB data (bottom-left) and jointly CMB and LSS data
  (bottom-right). The contours represent the subsequent 68\% and 95\%
  confidence regions. While SN1a data induce constraints that are quite loose,
  LSS and CMB constraints are much tighter and tend to favor a
  cosmological constant like scenario.\label{fig7}}
\end{figure*}
%----------------------------------------------------------

\section{Chaplygin gas likelihood analysis}
%============================================================
\label{sec5}

In order now to assess the viability of a GCG+CDM+baryon universe, we
turn to evaluate the probability (the posterior) of these models
given some current observations, namely SN1a, CMB and LSS probed
through galaxy survey. 

To study the posterior distribution, we use the Baye's theorem and
rewrite it as the product of the likelihood and the prior (we assume
the evidence is constant and thus ignore it). To probe
this posterior, we consequently compute both the likelihood and the prior at
various positions in the chosen restricted parameter space. This sampling
is conducted via the construction of a Monte Carlo Markov Chain
through the Metropolis-Hasting algorithm. Once converged, this chain
provides us with a collection of independant samples from the
posterior (see \cite{ChMe00,ChMe01,MCMC} for an introduction to this
technique in this context and \cite{ChGr95,Gil96} for general guidance). 

Our code uses some likelihood computation elements from the code
described in \cite{MCMC}, and relies on a version of the CAMB code 
\cite{CAMB} extended to include a Chaplygin gas component in order to calculate
CMB power spectra and matter power spectra. As input data, we considered the
apparent magnitudes of 51 Supernovae \cite{superP}, CMB data sets from COBE \cite{COBE}, MAXIMA
\cite{MAX}, BOOMERANG \cite{Boom}, and VSA \cite{VSA} and large scale
structure data from 2dF \cite{2dF}. We consider only flat models, \ie
$\Omega_{K}=0$ with scale invariant initial power spectrum, \ie $n_s=1$. We use stringent (Gaussian)
priors on $H_{0}$ using the HST Key Project results  $h = 0.72\pm
0.08$ \cite{HST}  and on $\omega_{b}=\Omega_{b}h^{2}=0.02 \pm 0.001$ using
BBN constraints \cite{BBN}.  

We normalize the matter power spectrum using $A_{s}$, the initial 
power spectrum normalization, and following \cite{Hamilton92}, we use
$\beta$ and $b_1$ to parameterize redshift-space distortions and 
(linear) bias respectively. The power spectrum is then related to the
transfer function $T(k)$ (computed with CAMB) by  
\beq 
P(k)=A_{s}\left(1+{2\beta\over 3}+{\beta^{2}\over   5}\right)b^{2}_1T(k)^{2}. \label{matpow}
\eeq 
In order to alleviate the natural degeneracy between $A_s$ and $b_1$
(as far as LSS constraints are concerned), we use the 2dF results
\cite{Peacock01,Verde02} to impose strong (Gaussian) priors on $\beta$ and $b_1$,
i.e. $\beta=0.54\pm 0.09, b_1 = 1.04 \pm 0.11$. 

Throughout this analysis, we ensured the chains' convergence by
generating and comparing several of them (typically containing $~10^5$ elements) and
by checking the so-called ``parameter mixing'' amid them. After several trials, we
choose the proposal density for each parameter to be a Gaussian whose
width is close to the final one and whose center is the last chain
values. This allows a full exploration of the parameter space. To
pick-up the next chain element, we allow only 1 to 3 directions (this
number is randomly chosen) to vary. This gives us an acceptance rate
around 25\%, a good target value for efficiency's sake
\cite{Gil96}. The first 4000 elements of the chain, prior to its
convergence, are thrown away and no extra thinning is applied
\cite{Gil96}.     

Once converged, the chains provide a fair sampling of the full
posterior distribution so that we can deduce easily from it all the
quantities of interest, \eg the (joint) marginalized distribution of any parameter(s).

As stated above, we are interested in finding the compatibility of a
Chaplygin gas + CDM + baryon universe with current data. For this we
vary only 8 parameters $\{h,\omega_{b},\omega_{cdm},\alpha,w_{0},b,\beta,A_{s}\}$
and impose the priors stated above. We allow a free proposal distribution for $\omega_c$, including $\omega_c=0$ consistent with a unified matter universe purely containing a GCG and baryons. This allows the full breadth of GCG roles (as both a dark matter and dark energy candidate) to be tested. Following section \ref{sec3}'s
discussion, we restrict ourselves to $0\le\alpha\le 1$ and $-1\le
w_0\le0$. In figure \ref{fig7} we plot the marginalized 
joint distribution of the $\alpha$,  $|w_{0}|$ parameters (which is in
this case just the joint number density of those parameters), as well as the 68\%
and 95\% confidence contours, considering separately SN1a, LSS, CMB
data sets, and also jointly CMB and LSS. Note that for visual purposes
only the displayed surface has been build by oversampling our samples
using cubic interpolation. This does not affect the quantitative
interpretation since the distributions turn out to be smooth. 

The interpretation of the contours is nicely consistent with the
theoretical prospects discussed above. First, the SN1a observations
(top-left panel) offer very light constraints on the GCG parameters,
since they are sensitive only to the background evolution. Any
$\alpha$ value appear viable, extending thus the obvious degeneracy
between $\Lambda$-CDM model and GCG models with $\alpha=0$ discussed
in section \ref{sec2} (\eg see figure \ref{fig1}). As soon as density perturbations are
considered, the constraints tighten drastically. For both
LSS and CMB, the isocontours are roughly centered on the $\alpha=0$, $w_{0}=-1$
model, that corresponds to the GCG acting like a $\Lambda$ term. This fact is
emphasized in the joint CMB + LSS analysis. Note that in the limit that $w_{0}$ tends
to $-1$ (we however impose $|w_{0}|<1$), the GCG component tends to
behave like a $\Lambda$ term, irrespective of the precise value of
$\alpha$, thus leading to the observed degeneracy in the $-w_{0}=1$
direction. 

The other varied parameters, \ie $h$, $\omega_{b}$ and $\omega_{cdm}$,
as well as the flatness imposed $\Omega_ch$, exhibit (joint)
distributions similar to those found in 
typical $\Lambda$-CDM model studies (see \eg \cite{MCMC}). This
leads us to the the main conclusion of this study: the current
data tends to favor ordinary $\Lambda$ theory. When
marginalized over all other parameters, we indeed find,
$\alpha<0.5,0.93$, and $w_{0}<-0.85,-0.8$, both respectively at
the 68\% and 95\% confidence level.    

\section{Conclusions}
%====================
\label{sec6}

We have investigated the effect of a Chaplygin gas matter component in
the universe's ingredients, to see if such a component is consistent 
with observations and whether it is a feasible alternative to CDM and $\Lambda$. 

Through inherent degeneracies  with $\Lambda$ in the background
evolution, the Chaplygin gas models have a good fit with SN1a
data. These degeneracies are not present, however, in the perturbation
evolution. In particular the growth/suppression of both GCG and CDM 
density perturbations proves distinctive when comparing against
large scale structure observations; this statement is valid, of course, for all cases (for all $\alpha$) except, naturally, $w_{0}-=-1$ which is identical to $\Lambda$, and has no perturbations. The GCG also introduces a number
of distinguishing differences from the $\Lambda$-CDM CMB spectrum
through altering the potential at last scattering, the ISW signature,
the equality scale, and the angular diameter distance to last
scattering. Combined, these differences provide a strong test for the
GCG scenario. 

We performed likelihood analyses using SN1a, CMB and LSS datasets and
found that the current data strongly prefers a $\Lambda$-like
dark energy component, with $\alpha<0. 5$ and $w_{0}<-0.85$ at the 68\%
level and with CDM as the preferred pressureless matter
component.%
 Note that, in comparison, the unified dark matter model, with $\Omega_{c}=0$, is highly disfavored by the data. 
This result is consistent, but considerably tightens, previous
 constraints from supernovae, CMB peak position and matter power
 spectrum shape parameter analyses
 (\cite{Kamen01}-\cite{Carturan02}). Our constraints can be recast in
 terms of the  `statefinder' parameters of \cite{Sahni02}, $r<1.20$
 and $s>-0.075$ at the 68\% level, thus greatly reducing the ability of a Chaplygin Gas
 to explain the `cosmic conundrum' problem as proposed in \cite{Gorini02}.

Our analysis assumed adiabatic perturbations for the Chaplygin gas; it
remains to be seen how enriching this model by considering
non-adiabatic perturbations, as mentioned in a paper presented after
the initial posting of this work \cite{Balakin03}, might alter the
analysis. 

On the basis of current observations however, Chaplygin gases, with
adiabatic perturbations at least, do not seem to provide a favored alternative
to scenarios involving CDM and a cosmological constant.
 
\vspace{1cm}
{\bf Acknowledgements} We would like to thank Elena Pierpaoli, David
Spergel, and Licia Verde for very helpful discussions in the course of
this work. RB and OD are supported by MAP and NASA ATP grant NAG5-7154 respectively.

\section{Appendix - Perturbation evolution for a general, adiabatic fluid}\label{app}
%=========================================================================

We use the basic background equation (time derivatives with
respect to $\tau$): 
\bea 
\dot{{\cal H}}+{\cal H}^{2}&=&{\ddot{a}\over a}={{\cal H}^{2}\over 2}\left[1-3\sum_{i}
(\Omega_{i}w_{i})\right] \label{ddotaoa} 
\eea
and  equation of state and speed of sound equations: \bea
 c_{s}^{2}&=&{{\delta P}\over
\delta{\rho}}={\dot{P}\over \dot{\rho}} \ \ \ (\mathrm{assuming\
adiabaticity})
\\ \dot{w}&=&{\dot{\rho}\over \rho}\left({\dot{P}\over \dot{\rho}}-{P\over \rho}\right)=-3{\cal
H}(1+w)(c_{s}^{2}-w) \eea

Following \cite{Ma95} the first order perturbation equations are
\bea
\label{deltaeq-app}\dot{\delta}&=&-\left(1+w\right)\left(\theta+{\dot{h}\over
  2}\right)-3\left(c_{s}^{2}-w\right){\cal H}\delta \\ 
\dot{\theta}&=&-\left(1-3c_{s}^{2}\right){\cal
H}\theta+{c_{s}^{2}\over (1+w)}k^{2}\delta .\eea
So that the second order equation in $\delta$ (differentiating
(\ref{deltaeq}) is given by
 \bea\label{deltaeq2-app}
\ddot{\delta}&=&-\left(1+w\right)\left(\dot{\theta}+{\ddot{h}\over
2}\right)-\dot{w}\left(\theta+{\dot{h}\over 2}\right) \nonumber
\\&&-3\left(c_{s}^{2}-w\right)(\dot{{\cal H}}\delta + {\cal
H}\dot{\delta})-3\left(\dot{c_{s}^{2}}-\dot{w}\right){\cal
H}\delta .\eea
We eliminate the time derivatives of the metric perturbations, $h$ and $\eta$,
using the perturbed Einstein equations
\bea  k^{2}\eta&-&{1\over 2}{\cal H}\dot{h}=-{1\over
2}a^{2}\delta\rho \\
\ddot{h}&+&2{\cal H}-2k^{2}\eta=-3a^{2}\delta P
\eea
which give
\bea(1+w){\ddot{h}\over 2} &=& {\cal
H}\dot{\delta}+(1+w){\cal H}\theta+3(c_{s}^{2}-w){\cal
H}^{2}\delta\nonumber \\ &&-(1+w){a^{2}\over 2}(\delta\rho+3\delta
P).
\eea

Collecting terms together we obtain the general evolution equation
for $\delta$ for any fluid with equation of state $w$ and speed of
sound $c_{s}^{}$, \bea \ddot{\delta}&=&-3c_{s}^{2}(1+w){\cal
H}\theta \nonumber -[1+6(c_{s}^{2}-w)]{\cal H}\dot{\delta}
-\left[c_{s}^{2}k^{2}+ \right.\nonumber \\ &&\left.
9(c_{s}^{2}-w)^{2}{\cal
H}^{2}+3\left(\dot{c_{s}^{2}}-\dot{w}\right){\cal
H}+3{\ddot{a}\over a}(c_{s}^{2}-w)\right]\delta \nonumber \\&&
+(1+w){a^{2}\over 2}(\delta\rho+3\delta P). \label{gen}\eea
Specializing to the Chaplygin gas in the matter dominated era\bea
{a^{2}\over 2}(\delta\rho+3\delta P)&\approx& {3{\cal H}^{2}\over
2}\left[\Omega_{ch}(1+3c_{s}^{2})\delta+\Omega_{c}\delta_{c}\right]
\\ {\ddot{a}\over a} &=&{{\cal
H}^{2}\over 2}(1-3\Omega_{ch}w)\\
c_{s}^{2}&=&-\alpha w \\
\dot{c_{s}^{2}}&=&-3\alpha{\cal H}w(1+w)(1+\alpha)\eea
we find \bea \ddot{\delta}&+&[1-6w(\alpha+1)]{\cal
H}\dot{\delta}-\left[\alpha w k^{2}+{3{\cal H}^{2}\over
2}\left\{\Omega_{ch}+\left(7+\Omega_{ch}\right.\right.\right.\nonumber
\\  &&\left.\left. \left.+(13-3\Omega_{ch})\alpha
+6\alpha^{2}\right)w
-3\Omega_{ch}(1+2\alpha)w^{2}\right\}\right]\delta \nonumber
\\&=&{3{\cal
H}^{2}\over 2}(1+w)\Omega_{c}\delta_{c}+3\alpha w(1+w){\cal
H}\theta .\eea

Similarly if one applies equation (\ref{gen}) to CDM with $w=c_{s}^{2}=0$, we find
\bea\ddot{\delta}_{c}&+&{\cal
H}\dot{\delta}_{c}-{3{\cal H}^{2}\over
2}\left[\Omega_{c}\delta_{c}+(1-3\alpha
w)\Omega_{ch}\delta\right]=0. \eea

\begin{thebibliography}{99}
%sec1
\bibitem{super1} P.M. Garnavich et al, \ApJL \textbf{493}, L53
(1998).

\bibitem{superP} S. Perlmutter et al, \ApJ \textbf{483}, 565 (1997); S. Perlmutter et al (The Supernova Cosmology Project), \Nature
 \textbf{391} 51 (1998); A.G. Riess et al, \ApJ \textbf{116} 1009
(1998); B.P. Schmidt, \ApJ \textbf{507} 46-63 (1998)

\bibitem{Peebles02} P. J. E. Peebles, B. Ratra, RMP (2003) in
press, astro-ph/0207347. 
\bibitem{Gibbons03} G. W. Gibbons, to appear in {\it Quantum and
Classical Gravity}, astro-ph/0301117.
\bibitem{Bilic01}N. Bili$\acute{c}$, G.~B. Tupper, R.~D. Viollier,
astro-ph/0111325.

\bibitem{Teg02} H. Sandvik, M. Tegmark, M. Zaldarriaga, I. Waga,
astro-ph/0212114.

\bibitem{Kamen01}A. Yu. Kamenshchik, U. Moschella, V. Pasquier, \PL B. \textbf{511} 265 (2001),gr-qc/0103004.

\bibitem{Bento02}M.~C. Bento, O. Bertolami, A.~A. Sen,
\PR D \textbf{66} 043507 (2002), astro-ph/0202064; astro-ph/0210375; astro-ph/0210468.
\bibitem{Fabris02} J.~C. Fabris, S.~V.~B. Goncalves, P.~E.
deSouza {\it Gen. Rel. Grav.} \textbf{34} 2111 (2002), astro-ph/0203441; {\it Gen. Rel. Grav.} \textbf{34} 53 (2002), astro-ph/0103083; astro-ph/0207430.
\bibitem{Avelino02} P. P. Avelino \etal, \PR D \textbf{67} 023511 (2002), astro-ph/0208528.
\bibitem{Carturan02} D. Carturan, F. Finelli, astro-ph/0211626.
\bibitem{Gonz02} P. F. Gonz$\acute{a}$lez-D$\acute{i}$az, astro-ph/0212414.


\bibitem{Deffayet02} C. Deffayet,S. Landau, J. Raux, M.
Zaldarriaga, P. Astier, \PR D \textbf{66} 024019 (2002), astro-ph/0201164.

\bibitem{Podolsky02}D. Podolsky, {\it Astron.Lett.} \textbf{28} 434 (2002), gr-qc/0203010.

%sec3
\bibitem{Ma95}   C-P. Ma, E. Bertschinger, \ApJ \textbf{455} 7 (1995),
astro-ph/9506072.

%sec4
\bibitem{Seljak96} U. Seljak, M. Zaldarriaga \ApJ \textbf{469} 437 (1996),
astro-ph/0006436.
\bibitem{Bardeen} J.M. Bardeen \PR D, \textbf{22} 1882 (1980) 
\bibitem{Bean01} R. Bean, A. Melchiorri \PR D \textbf{65} 041302 (2002),
    astro-ph/0110472.
\bibitem{Hu01} W. Hu, M. Fukugita, M., Zaldarriaga, M. Tegmark \ApJ
  \textbf{549} 669 (2001).

%sec5
\bibitem{ChMe00} N. Christensen, R. Meyer, astro-ph/0006401
\bibitem{ChMe01} N. Christensen, R. Meyer, L. Knox, B. Luey, {\it Classical and Quantum Gravity}  \textbf{18} 2677 (2001), astro-ph/0103134
\bibitem{MCMC} A. Lewis, S. Bridle, astro-ph/0205436.
\bibitem{Gil96} Eds W.R. Gilks, S. Richardson, D.J. Spiegelhalter, {\it Markov
  Chain in Practice}, Chapman \& Hall, 1996

\bibitem{ChGr95} S. Chib \& E. Greenberg, {\it The American Statistician} \textbf{49} 4 (1995).
\bibitem{CAMB}A. Lewis, A. Challinor, A. Lasenby, \ApJ \textbf{538} 473
(2000).
\bibitem{COBE}G. Smoot \etal , \ApJ \textbf{386} L1 (1992), C. Bennett \etal \ApJ \textbf{464} L1 (1996).
\bibitem{MAX}S. Hanany \etal, \ApJ Lett. \textbf{545} 5 (2000), astro-ph/0005123.
\bibitem{Boom} C.B. Netterfield \etal, \ApJ \textbf{571} 604 (2002), astro-ph/0104460.
\bibitem{VSA} P. F. Scott \etal, astro-ph/0205380.
\bibitem{2dF} O. Lahav \etal, {\it MNRAS} \textbf{333} 961L (2002), astro-ph/0112162.
\bibitem{HST} W. L. Freedman  \etal, \ApJ \textbf{553} 47 (2001)
, astro-ph/0012376.
\bibitem{BBN} S. Burles, K.M. Nollett, \& M. S. Turner, \ApJ
\textbf{552} L1 (2001), astro-ph/0010171.
\bibitem{Hamilton92} A.J.S. Hamilton, \ApJ \textbf{385},l5 (1992).
\bibitem{Peacock01} J. A. Peakcock \etal, \Nature
\textbf{401} 169, astro-ph/0103143.
\bibitem{Verde02}L. Verde \etal, {\it MNRAS} \textbf{335} 432,
astro-ph/0112161.
\bibitem{Sahni02}V. Sahni, T. D. Saini, A. A. Starobinsky, U. Alam, to appear in {\it JETP Lett},astro-ph/0201498. 
\bibitem{Gorini02} V. Gorini, A. Kamenshchik, U. Moschella, astro-ph/0209395.
\bibitem{Balakin03} A. B. Balakin, D. Pav\'{o}n, D. Schwarz, W. Zimdahl, astro-ph/0302150.

\end{thebibliography}
\end{document}